\documentclass[a4,12pt]{article}

\begin{document}

\begin{center}
{\Large\bf Broken $S^{}_3$ Flavor Symmetry of Leptons
and Quarks: Mass Spectra and Flavor Mixing Patterns}
\end{center}

\vspace{0.3cm}
\begin{center}
{\bf Zhi-zhong Xing}$^{~\rm a}$ \footnote{E-mail:
xingzz@ihep.ac.cn}, ~{\bf Deshan Yang}$^{~\bf b}$ \footnote{E-mail:
yangds@gucas.ac.cn},
~{\bf Shun Zhou}$^{~\rm c}$ \footnote{E-mail: zhoush@mppmu.mpg.de} \\
$^{\rm a}$ Institute of High Energy Physics, Chinese Academy of
Sciences, Beijing 100049, China \\
$^{\rm b}$ Graduate University of
Chinese Academy of Sciences, Beijing 100049, China \\
$^{\rm c}$ Max-Planck-Institut f$\rm\ddot{u}$r Physik
(Werner-Heisenberg-Institut), 80805 Munich, Germany
\end{center}

\vspace{2.5cm}

\begin{abstract}
We apply the discrete $S^{}_3$ flavor symmetry to both lepton and
quark sectors of the standard model extended by introducing one
Higgs triplet and realizing the type-II seesaw mechanism for finite
neutrino masses. The resultant mass matrices of charged leptons
($M^{}_l$), neutrinos ($M^{}_\nu$), up-type quarks ($M^{}_{\rm u}$)
and down-type quarks ($M^{}_{\rm d}$) have a universal form
consisting of two terms: one is proportional to the identity matrix
$I$ and the other is proportional to the democracy matrix $D$. We
argue that the textures of $M^{}_l$, $M^{}_{\rm u}$ and $M^{}_{\rm
d}$ are dominated by the $D$ term, while that of $M^{}_\nu$ is
dominated by the $I$ term. This hypothesis implies a near mass
degeneracy of three neutrinos and can naturally explain why the mass
matrices of charged fermions are strongly hierarchical, why the
quark mixing matrix is close to $I$ and why the lepton mixing matrix
contains two large angles. We discuss a rather simple perturbation
ansatz to break the $S^{}_3$ symmetry and obtain more realistic mass
spectra of leptons and quarks as well as their flavor mixing
patterns. We stress that the $I$ term, which used to be ignored from
$M^{}_l$, $M^{}_{\rm u}$ and $M^{}_{\rm d}$, is actually important
because it can significantly modify the smallest lepton flavor
mixing angle $\theta^{}_{13}$ or three quark flavor mixing angles.
\end{abstract}

\newpage

\framebox{\large\bf 1} ~ Flavor (or family) symmetries belong to one
of a few promising approaches toward deeper understanding of the
observed mass spectra and flavor mixing patterns of leptons and
quarks \cite{Harari,FN,FX00}. Among a number of interesting discrete
flavor symmetries discussed in the literature \cite{Review}, the
$S^{}_3$ symmetry should be the simplest one and has been used to
interpret the mass hierarchies of charged fermions \cite{PDG} and
predict the well-known ``democratic" \cite{FX96} and
``tri-bimaximal" \cite{TB} neutrino mixing scenarios. A particular
merit of the $S^{}_3$ flavor symmetry is that it requires the mass
matrices of three charged leptons ($M^{}_l$), three neutrinos
($M^{}_\nu$), three up-type quarks ($M^{}_{\rm u}$) and three
down-type quarks ($M^{}_{\rm d}$) to have a universal form $M^{}_x =
\xi^{}_x I + \zeta^{}_x D$ (for $x= l, \nu, {\rm u}, {\rm d}$),
where
\begin{eqnarray}
I \; = \; \left(\matrix{1 & 0 & 0 \cr 0 & 1 & 0 \cr 0 & 0 &
1}\right) \; , ~~~ D \; = \; \left(\matrix{1 & 1 & 1 \cr 1 & 1 & 1
\cr 1 & 1 & 1} \right) \;
\end{eqnarray}
stand respectively for the identity and democracy matrices. Hence
one may explore the flavor structures of leptons and quarks, both
similarities and differences of their mass spectra and flavor mixing
patterns, in a more or less parallel way.

The above point can be made clear if we extend the standard
electroweak model by introducing one Higgs triplet and realizing the
type-II seesaw mechanism \cite{T2} to generate finite neutrino
masses. In this case the gauge-invariant Lagrangian relevant to
quark and lepton masses reads
\begin{eqnarray}
-{\cal L}^{}_{\rm Q + L} & = & \overline{Q^{}_{\rm L}} Y^{}_{\rm u}
\tilde{H} U^{}_{\rm R} + \overline{Q^{}_{\rm L}} Y^{}_{\rm d} H
D^{}_{\rm R} + \overline{\ell^{}_{\rm L}} Y^{}_l H E^{}_{\rm R} +
\frac{1}{2} \overline{\ell^{}_{\rm L}}
Y^{}_\Delta \Delta i\sigma^{}_2 \ell^c_{\rm L} ~~~~ \nonumber \\
&& - \lambda^{}_\Delta
M^{}_\Delta H^T i\sigma^{}_2 \Delta H + {\rm h.c.} \; ,
\end{eqnarray}
where $Q^{}_{\rm L}$ and $\ell^{}_{\rm L}$ are the left-handed
doublets of quarks and leptons, $U^{}_{\rm R}, D^{}_{\rm R},
E^{}_{\rm R}$ are the right-handed singlets of up-type quarks,
down-type quarks and charged leptons, $H$ and $\Delta$ stand
respectively for the Higgs doublet and triplet. The last term of
${\cal L}^{}_{\rm Q+L}$ is lepton-number-violating, and thus its
dimensionless coefficient $\lambda^{}_\Delta$ is naturally small
\cite{Hooft}. Once the neutral components of $H$ and $\Delta$
acquire their vacuum expectation values, the $SU(2)^{}_{\rm L}
\times U(1)^{}_{\rm Y}$ gauge symmetry is spontaneously broken such
that all of the quarks and leptons become massive:
\begin{eqnarray}
-{\cal L}^{}_{\rm mass} \; = \; \overline{U^{}_{\rm L}} M^{}_{\rm u}
U^{}_{\rm R} + \overline{D^{}_{\rm L}} M^{}_{\rm d} D^{}_{\rm R} +
\overline{E^{}_{\rm L}} M^{}_l E^{}_{\rm R} + \frac{1}{2}
\overline{\nu^{}_{\rm L}} M^{}_\nu \nu^c_{\rm L} + {\rm h.c.} \; ,
\end{eqnarray}
where the charged-fermion mass matrices are given by $M^{}_x \equiv
Y^{}_x \langle H\rangle$ (for $x = {\rm u}, {\rm d}, l$) with
$\langle H\rangle \approx 174 ~{\rm GeV}$, and the Majorana neutrino
mass matrix is given by $M^{}_\nu = Y^{}_\Delta \langle \Delta
\rangle$ with $\langle \Delta \rangle = 2 \lambda^{}_\Delta \langle
H\rangle^2/M^{}_\Delta$. The mass scale of the Higgs triplet
$M^{}_\Delta$ is expected to be much higher than the electroweak
scale characterized by $\langle H\rangle$, so the tiny mass scale of
$M^{}_\nu$ is apparently attributed to the smallness of
$\lambda^{}_\Delta$ and the largeness of $M^{}_\Delta$. Current
experimental data on the electroweak $\rho$ parameter require
$\langle \Delta \rangle < 2~{\rm GeV}$ \cite{PDG}. Now we apply the
non-Abelian discrete symmetry $S^{}_3$, a permutation group of three
objects, to the lepton and quark sectors. It contains six elements,
\begin{eqnarray}
S^{(123)} & = & \left(\matrix{1 & 0 & 0 \cr 0 & 1 & 0 \cr 0 & 0 &
1}\right) \; , ~~~ S^{(213)} \; = \; \left(\matrix{0 & 1 & 0 \cr
1 & 0 & 0 \cr 0 & 0 & 1}\right) \; , \nonumber \\
S^{(132)} & = & \left(\matrix{1 & 0 & 0 \cr 0 & 0 & 1 \cr 0 & 1 &
0}\right) \; , ~~~ S^{(321)} \; = \; \left(\matrix{0 & 0 & 1 \cr
0 & 1 & 0 \cr 1 & 0 & 0}\right) \; , \nonumber \\
S^{(312)} & = & \left(\matrix{0 & 0 & 1 \cr 1 & 0 & 0 \cr 0 & 1 &
0}\right) \; , ~~~ S^{(231)} \; = \; \left(\matrix{0 & 1 & 0 \cr 0 &
0 & 1 \cr 1 & 0 & 0}\right) \; ,
\end{eqnarray}
which fall into three conjugacy classes. Allowing the lepton and
quark fields to transform as $Q^{}_{\rm L} \to S^{(ijk)} Q^{}_{\rm
L}$, $\ell^{}_{\rm L} \to S^{(ijk)} \ell^{}_{\rm L}$, $U^{}_{\rm R}
\to S^{(ijk)} U^{}_{\rm R}$, $D^{}_{\rm R} \to S^{(ijk)} D^{}_{\rm
R}$ and $E^{}_{\rm R} \to S^{(ijk)} E^{}_{\rm R}$, one may easily
find that the gauge interactions (or kinetic terms) of leptons and
quarks are invariant. The Yukawa interactions of leptons and quarks
in Eq. (2) are also invariant under the same transformations if the
following commutation conditions are satisfied:
\begin{equation}
\left[Y^{}_{\rm u} ~, S^{(ijk)}\right] \; = \; \left[Y^{}_{\rm d} ~,
S^{(ijk)}\right] \; = \; \left[Y^{}_l ~, S^{(ijk)}\right] \; = \;
\left[Y^{}_\Delta ~, S^{(ijk)}\right] \; = \; {\bf 0} \; .
\end{equation}
The mass matrices $M^{}_x$ (for $x= {\rm u}, {\rm d}, l, \nu$) must
equivalently satisfy $\left[M^{}_x ~, S^{(ijk)}\right] ={\bf 0}$,
implying that they can only take a universal form $M^{}_x = \xi^{}_x
I + \zeta^{}_x D$ in the $S^{}_3$ symmetry limit with $I =
S^{(123)}$ and $D = S^{(213)} + S^{(132)} + S^{(321)} = S^{(123)} +
S^{(312)} + S^{(231)}$ \cite{Schechter}. Because an orthogonal
matrix with two large rotation angles is needed to diagonalize the
democracy matrix $D$, such as
\begin{eqnarray}
U^T_0 D U^{}_0 & = & \left( \matrix{ 0 & 0 & 0 \cr 0 & 3 & 0 \cr 0 &
0 & 0 \cr} \right) \; \equiv \; H^{}_2 \; , ~~~ U^{}_0 \; =\;
\frac{1}{\sqrt{6}} \left( \matrix{ -2 & \sqrt{2} & 0 \cr 1 &
\sqrt{2} & \sqrt{3} \cr 1 & \sqrt{2} & -\sqrt{3} \cr} \right) \; ;
\nonumber \\
V^T_0 D V^{}_0 & = & \left( \matrix{ 0 & 0 & 0 \cr 0 & 0 & 0 \cr 0 &
0 & 3 \cr} \right) \; \equiv \; H^{}_3 \; , ~~~ V^{}_0 \; =\;
\frac{1}{\sqrt{6}} \left( \matrix{ \sqrt{3} & 1 & \sqrt{2} \cr
-\sqrt{3} & 1 & \sqrt{2} \cr 0 & -2 & \sqrt{2} \cr} \right) \; ,
\end{eqnarray}
it is potentially possible to explain the large solar and
atmospheric neutrino mixing angles. One can see that $V^T_0$
corresponds to the democratic mixing pattern \cite{FX96} while
$U^{}_0$ is just the tri-bimaximal mixing pattern \cite{TB}. Since
$H^{}_3$ contains a non-vanishing (dominant) matrix element in the
(3,3) position, it is consistent with the observed mass hierarchies
of charged fermions (i.e., $m^{}_t \gg m^{}_c \gg m^{}_u$, $m^{}_b
\gg m^{}_s \gg m^{}_d$ and $m^{}_\tau \gg m^{}_\mu \gg m^{}_e$) in
the symmetry limit. This observation means that $\xi^{}_x \ll
\zeta^{}_x$ (for $x = {\rm u}, {\rm d}, l$) is likely to hold for
charged fermions, while $\xi^{}_\nu \gg \zeta^{}_\nu$ is more
reasonable to generate two large lepton flavor mixing angles
together with a nearly degenerate neutrino mass spectrum
\cite{FX96,Tanimoto,Yanagida,FX04,RX04}. Of course, proper perturbations to
the lepton or quark mass matrices are necessary \cite{FX00,Koide} in
order to stabilize the dominant term of the quark or lepton flavor
mixing matrix and produce appreciable CP violation in the lepton or
quark sector.

In this letter we stress that the $I$ term of $M^{}_x$, which used
to be ignored for charged fermions (i.e., $\xi^{}_l = \xi^{}_{\rm u}
= \xi^{}_{\rm d} =0$ was often assumed or dictated by the $S^{}_{3
\rm L} \times S^{}_{3 \rm R}$ symmetry), is actually important. On
the one hand, this term is present in the $S^{}_3$ symmetry limit
and thus has no good reason to be switched off. On the other hand,
small $\xi^{}_l$ can significantly modify the smallest lepton flavor
mixing angle $\theta^{}_{13}$, while small $\xi^{}_{\rm u}$ and
$\xi^{}_{\rm d}$ can remarkably modify three quark flavor mixing
angles. We illustrate our points by discussing a rather simple
perturbation ansatz to explicitly break the $S^{}_3$ symmetry and
obtain more realistic mass spectra of leptons and quarks as well as
their flavor mixing patterns.

\vspace{0.3cm}

\framebox{\large\bf 2} ~ Let us first consider the lepton sector.
Based on the $S^{}_3$ symmetry discussed above, the mass matrices of
charged leptons and neutrinos can be written as
\begin{eqnarray}
M^{}_l &=& \frac{c^{}_l}{3} \left[ \left(\matrix{1 & 1 & 1 \cr 1 & 1
& 1 \cr 1 & 1 & 1}\right) + r^{}_l \left(\matrix{1 & 0 & 0 \cr 0 &
1 & 0 \cr 0 & 0 & 1}\right)\right] + \Delta M^{}_l \; , \nonumber \\
M^{}_\nu &=& c^{}_\nu \left[ \left(\matrix{1 & 0 & 0 \cr 0 & 1 & 0
\cr 0 & 0 & 1}\right) + r^{}_\nu \left(\matrix{1 & 1 & 1 \cr 1 & 1 &
1 \cr 1 & 1 & 1}\right) \right] + \Delta M^{}_\nu \; ,
\end{eqnarray}
where $c^{}_x >0$, $r^{}_x$ is real and $|r^{}_x| \ll 1$ holds (for
$x = l$ and $\nu$), and the explicit symmetry-breaking terms $\Delta
M^{}_l$ and $\Delta M^{}_\nu$ are assumed to take the diagonal forms
\cite{FX96,FX04}
\begin{eqnarray}
\Delta M^{}_l \; = \; \frac{c^{}_l}{3} \left(\matrix{-i\delta^{}_l &
0 & 0 \cr 0 & +i\delta^{}_l & 0 \cr 0 & 0 & \varepsilon^{}_l}\right)
\; , ~~~ \Delta M^{}_\nu \; = \; c^{}_\nu \left(\matrix{-
\delta^{}_\nu & 0 & 0 \cr 0 & + \delta^{}_\nu & 0 \cr 0 & 0 &
\varepsilon^{}_\nu}\right) \;
\end{eqnarray}
with $0 < \delta^{}_x \ll \varepsilon^{}_x \ll 1$ (for $x = l$ and
$\nu$). To diagonalize $M^{}_l$, one may transform it into the
following hierarchical texture:
\begin{eqnarray}
M^\prime_l \; \equiv \; V^T_0 M^{}_l V^{}_0 \; = \;
\frac{c^{}_l}{9\sqrt{3}} \left(\matrix{3\sqrt{3} r^{}_l & -i 3
\delta^{}_l & -i 3\sqrt{2} \delta^{}_l \cr -i 3 \delta^{}_l &
\sqrt{3} \left(3r^{}_l + 2 \varepsilon^{}_l\right) & -\sqrt{6}
\varepsilon^{}_l \cr -i 3\sqrt{2} \delta^{}_l & -\sqrt{6}
\varepsilon^{}_l & \sqrt{3} \left(9 + 3r^{}_l + \varepsilon^{}_l
\right) \cr} \right) \; .
\end{eqnarray}
where $V^{}_0$ has been given in Eq. (6). In the assumption of
$|r^{}_l| \ll \varepsilon^{}_l$, we can diagonalize $M^\prime_l$ and
obtain the masses of three charged leptons to a good degree of
accuracy:
\begin{eqnarray}
m^{}_\tau \; \approx \; c^{}_l \left(1 + \frac{\varepsilon^{}_l}{9}
+ \frac{r^{}_l}{3}\right) \; , ~~~ m^{}_\mu \; \approx \; c^{}_l
\left(\frac{2\varepsilon^{}_l}{9} + \frac{r^{}_l}{3}\right) \; , ~~~
m^{}_e \; \approx \; c^{}_l \left|
\frac{\delta^2_l}{6\varepsilon^{}_l} + \frac{r^{}_l}{3}\right| \; .
\end{eqnarray}
Defining $m^{}_0 \equiv c^{}_l r^{}_l/3$,
we have $|m^{}_0| < m^{}_\mu$.
Then $\varepsilon^{}_l$ and $\delta^{}_l$ are given by
\begin{eqnarray}
\varepsilon^{}_l \; \approx \; \frac{9}{2} \frac{m^{}_\mu -
m^{}_0}{m^{}_\tau - m^{}_0} \; , ~~~ \delta^{}_l \; \approx \;
\frac{2\varepsilon^{}_l}{\sqrt{3}} \frac{\sqrt{|m^{}_e -
|m^{}_0||}}{\sqrt{m^{}_\mu - m^{}_0}} \; .
\end{eqnarray}
As a consequence of Eq. (10), the constraint $m^{}_0 < m^{}_e$
should be satisfied for $m^{}_0 > 0$. The unitary matrix used to
diagonalize $M^{}_l$ (i.e., $V^\dagger_l M^{}_l V^*_l = {\rm
Diag}\{m^{}_e, m^{}_\mu, m^{}_\tau\}$) is found to deviate from
$V^{}_0$ at the level of ${\cal O}(\varepsilon^{}_l)$ and ${\cal
O}(\delta^{}_l)$:
\begin{eqnarray}
V^{}_l \; \approx \; V^{}_0 - \frac{i}{\sqrt{6}} \frac{\sqrt{|m^{}_e
- |m^{}_0||}}{\sqrt{m^{}_\mu - m^{}_0}} \left(\matrix{ 1 & \sqrt{3}
& 0 \cr 1 & -\sqrt{3} & 0 \cr -2 & 0 & 0} \right) +
\frac{1}{2\sqrt{3}} \frac{m^{}_\mu - m^{}_0}{m^{}_\tau - m^{}_0}
\left(\matrix{ 0 & \sqrt{2} & -1 \cr 0 & \sqrt{2} & -1 \cr 0 &
\sqrt{2} & 2} \right) \; .
\end{eqnarray}
On the other hand, the unitary matrix used to diagonalize the
Majorana neutrino mass matrix $M^{}_\nu$ (i.e., $V^\dagger_\nu
M^{}_\nu V^*_\nu = {\rm Diag}\{m^{}_1, m^{}_2, m^{}_3\}$) is
approximately given by \cite{FX04}
\begin{eqnarray}
V^{}_\nu \; \approx \; \frac{1}{\varepsilon^{}_\nu} \left(
\matrix{\varepsilon^{}_\nu c^{}_\theta & \varepsilon^{}_\nu
s^{}_\theta & r^{}_\nu \cr -\varepsilon^{}_\nu s^{}_\theta &
\varepsilon^{}_\nu c^{}_\theta & r^{}_\nu \cr r^{}_\nu
\left(s^{}_\theta - c^{}_\theta \right) & -r^{}_\nu
\left(s^{}_\theta + c^{}_\theta \right) & \varepsilon^{}_\nu}\right)
\; ,
\end{eqnarray}
where $c^{}_\theta \equiv \cos\theta$ and $s^{}_\theta \equiv
\sin\theta$ with $\tan 2\theta \equiv r^{}_\nu/\delta^{}_\nu$. Three
neutrino mass eigenvalues are
\begin{eqnarray}
m^{}_3 & \approx & c^{}_\nu \left(1 + r^{}_\nu + \varepsilon^{}_\nu \right) \; ,
\nonumber \\
m^{}_2 & \approx & c^{}_\nu \left(1 + r^{}_\nu + \sqrt{r^2_\nu +
\delta^2_\nu} \right) \; , \nonumber \\
m^{}_1 & \approx & c^{}_\nu \left(1 + r^{}_\nu - \sqrt{r^2_\nu +
\delta^2_\nu} \right) \; .
\end{eqnarray}
Therefore, $\Delta m^2_{21} \approx 4c^2_\nu \sqrt{r^2_\nu +
\delta^2_\nu}$ and $\Delta m^2_{32} \approx 2c^2_\nu
\varepsilon^{}_\nu$, from which $r^{}_\nu/\varepsilon^{}_\nu \approx
c^{}_\theta s^{}_\theta \Delta m^2_{21}/\Delta m^2_{32}$ can be
obtained. The $3\times 3$ lepton flavor mixing matrix is defined as
$V^{}_{\rm MNS} = V^\dagger_l V^{}_\nu$ \cite{MNS}. A
straightforward calculation leads us to
\begin{eqnarray}
V^{}_{\rm MNS} & \approx & \frac{1}{\sqrt{6}} \left ( \matrix{
\sqrt{3} \left ( c^{}_\theta + s^{}_\theta \right ) & -\sqrt{3}
\left ( c^{}_\theta - s^{}_\theta \right ) & 0 \cr \left (
c^{}_\theta - s^{}_\theta \right ) & \left ( c^{}_\theta +
s^{}_\theta \right ) & -2 \cr \sqrt{2} \left ( c^{}_\theta -
s^{}_\theta \right ) & \sqrt{2} \left ( c^{}_\theta + s^{}_\theta
\right ) & \sqrt{2} \cr} \right)
\nonumber \\
& & + \frac{i}{\sqrt{6}} \frac{\sqrt{|m^{}_e -
|m^{}_0||}}{\sqrt{m^{}_\mu - m^{}_0}} \left ( \matrix{ \left (
c^{}_\theta - s^{}_\theta \right ) & \left ( c^{}_\theta +
s^{}_\theta \right ) & -2 \cr \sqrt{3} \left ( c^{}_\theta +
s^{}_\theta \right ) & -\sqrt{3} \left ( c^{}_\theta - s^{}_\theta
\right )  & 0 \cr 0 & 0 & 0 \cr} \right )
\nonumber \\
& & + \frac{1}{2\sqrt{3}} \frac{m^{}_\mu - m^{}_0}{m^{}_\tau -
m^{}_0} \left ( \matrix{ 0 & 0 & 0 \cr \sqrt{2} \left ( c^{}_\theta
- s^{}_\theta \right ) & \sqrt{2} \left ( c^{}_\theta + s^{}_\theta
\right ) & \sqrt{2} \cr - \left ( c^{}_\theta - s^{}_\theta \right )
& - \left ( c^{}_\theta + s^{}_\theta \right ) & 2 \cr} \right )
\nonumber \\
&  & + \frac{1}{\sqrt{6}} \frac{r_\nu}{\varepsilon_\nu} \left (
\matrix{ 0 & 0 & 0 \cr 2 \left ( c^{}_\theta - s^{}_\theta \right )
& 2 \left ( c^{}_\theta + s^{}_\theta \right ) & 2 \cr -\sqrt{2}
\left ( c^{}_\theta - s^{}_\theta \right ) & -\sqrt{2} \left (
c^{}_\theta + s^{}_\theta \right ) & 2\sqrt{2} \cr} \right ) \; .
\end{eqnarray}
Comparing this result with the standard parametrization of
$V^{}_{\rm MNS}$ advocated in Ref. \cite{PDG} (see also Ref.
\cite{FX01}), we immediately arrive at three flavor mixing angles
\begin{eqnarray}
\theta^{}_{12} & \approx & \frac{\pi}{4} - \theta = \frac{\pi}{4} -
\frac{1}{2} \arctan \left(\frac{r^{}_\nu}{\delta^{}_\nu}\right)  \;
, \nonumber \\
\theta^{}_{13} & \approx & \arcsin \left( \frac{2}{\sqrt{6}}
\frac{\sqrt{|m^{}_e - |m^{}_0||}}{\sqrt{m^{}_\mu - m^{}_0}} \right)
\;, \nonumber \\
\theta^{}_{23} & \approx & \frac{1}{2} \arcsin \left[
\frac{2\sqrt{2}}{3} \left( 1 + \frac{1}{2} \frac{m^{}_\mu -
m^{}_0}{m^{}_\tau - m^{}_0} + \frac{r^{}_\nu}{\varepsilon^{}_\nu}
\right) \right] \; ,
\end{eqnarray}
together with the Dirac CP-violating phase $\delta \approx \pi/2$.
Two Majorana CP-violating phases are trivial in this scenario, and
thus the effective mass of the neutrinoless double-beta decay is
simply given by $\langle m\rangle^{}_{ee} \approx c^{}_\nu$.
Different from Ref. \cite{FX04}, here both $\theta^{}_{13}$ and
$\theta^{}_{23}$ get modified because of the non-vanishing $m^{}_0 =
c^{}_l r^{}_l/3$ which signifies an important contribution from the
$I$ term in the $S^{}_3$ symmetry limit. Some discussions are in
order.
\begin{itemize}
\item    A global analysis of current neutrino oscillation data
yields $\theta^{}_{12} \approx 34.5^\circ$ \cite{Maltoni}, implying
$\theta \approx 10.5^\circ$ or equivalently $r^{}_\nu/\delta^{}_\nu
\approx 0.38$. We see that the size of $\theta$ is smaller than the
Cabibbo angle of quark flavor mixing (i.e., $\theta^{}_{\rm C}
\approx 13^\circ$  \cite{PDG}), so the so-called quark-lepton
complementarity relation $\theta^{}_{12} + \theta^{}_{\rm C} \approx
\pi/4$ becomes less favored than before. In addition, we obtain
$r^{}_\nu/\varepsilon^{}_\nu \approx 5.7 \times 10^{-3}$ from
$r^{}_\nu/\varepsilon^{}_\nu \approx c^{}_\theta s^{}_\theta \Delta
m^2_{21}/\Delta m^2_{32}$ with the typical inputs $\Delta m^2_{21}
\approx 7.6 \times 10^{-5} ~{\rm eV}^2$ and $\Delta m^2_{32} \approx
2.4 \times 10^{-3} ~{\rm eV}^2$ \cite{Maltoni}.

\item    If $|m^{}_0|$ is significantly larger than $m^{}_e$, the smallest
neutrino mixing angle $\theta^{}_{13}$ will be increased. In this
case, the magnitude of $\theta^{}_{13}$ is sensitive to that of
$m^{}_0$ , and thus the latter can get constrained from the present
experimental data. With the help of Eq. (16), we obtain
\begin{eqnarray}
m^{}_0 \; \approx \; - \frac{2 m^{}_e + 3 m^{}_\mu
\sin^2\theta^{}_{13}}{2 - 3\sin^2\theta^{}_{13}} \approx -\left(
m^{}_e + \frac{3}{2} m^{}_\mu \sin^2\theta^{}_{13}\right) \; ,
\end{eqnarray}
for $m^{}_0 < 0$ and $|m^{}_0| > m^{}_e$. Given $m^{}_e \approx
0.4866$ MeV and $m^{}_\mu \approx 102.718$ MeV at the electroweak
scale \cite{XZZ08}, the upper bound $\theta^{}_{13} \leq 12^\circ$
\cite{Maltoni} leads us to $-14.7 m^{}_e \leq m^{}_0 < - m^{}_e$ at
the same energy scale. In view of $r^{}_l = 3m^{}_0/c^{}_l \approx
3m^{}_0/m^{}_\tau$ together with $m^{}_\tau \approx 1746.24$ MeV at
the electroweak scale \cite{XZZ08}, we find $-1.2 \times 10^{-2}
\leq r^{}_l < -8.4 \times 10^{-4}$. So the magnitude of $r^{}_l$ is
strongly suppressed. Of course, we have $\theta^{}_{13} \approx
0^\circ$ for $|m^{}_0| \approx m^{}_e$ and $\theta^{}_{13} \approx
\arcsin\sqrt{2m^{}_e/(3m^{}_\mu)} \approx 3.2^\circ$ for $m^{}_0
\approx 0$. Once the value of $\theta^{}_{13}$ is experimentally
fixed, one may then be able to determine the important
$S^{}_3$-symmetry parameter $m^{}_0$.

\item    Because of $|m^{}_0|/m^{}_\mu < 15 m^{}_e/m^{}_\mu \approx 0.071$
at the electroweak scale, the $m^{}_0$-induced correction to
$\theta^{}_{23}$ in Eq. (16) is apparently negligible. Thus
$\varepsilon^{}_l \approx 9 m^{}_\mu/(2 m^{}_\tau) \approx 0.26$ and
$\delta^{}_l \approx \sqrt{2} \varepsilon^{}_l \sin\theta^{}_{13}
\leq 7.6 \times 10^{-2}$. The smallness of
$r^{}_\nu/\varepsilon^{}_\nu$ also makes its contribution to
$\theta^{}_{23}$ negligible. Taking $m^{}_\mu/m^{}_\tau \approx
0.0588$ \cite{XZZ08}, we obtain $\theta^{}_{23} \approx 38^\circ$.
This result, which certainly depends on the assumed perturbation
forms of $\Delta M^{}_l$ and $\Delta M^{}_\nu$, is a bit lower than
the best-fit value of $\theta^{}_{23}$ (i.e., $\theta^{}_{23} =
42.8^\circ$ \cite{Maltoni}) but lies in the $3\sigma$ interval of
$\theta^{}_{23}$ (i.e., $\theta^{}_{23} =
42.8^{+10.7^\circ}_{-7.3^\circ}$ \cite{Maltoni}).
\end{itemize}
Note that the magnitude of $r^{}_\nu$ can be determined only after
the absolute mass scale of three neutrinos (i.e., $c^{}_\nu$) is
known. Taking $c^{}_\nu \approx 0.1$ eV for example, we find
$\varepsilon^{}_\nu \approx \Delta m^2_{32}/(2c^2_\nu) \approx 0.12$
and therefore $r^{}_\nu \approx 6.8 \times 10^{-4}$. The smallness
of both $r^{}_l$ and $r^{}_\nu$ is consistent with our original
expectations. Although $r^{}_l$ and $r^{}_\nu$ are two free
parameters, they are intrinsical in the $S^{}_3$ flavor symmetry and
hence should be taken into account. We have demonstrated that both
of them are phenomenologically important in understanding the lepton
mass hierarchy and the flavor mixing pattern.

\vspace{0.3cm}

\framebox{\large\bf 3} ~ We proceed to look at the quark sector.
Based on the $S^{}_3$ flavor symmetry, the mass matrices of up- and
down-type quarks are written as
\begin{eqnarray}
M^{}_{\rm u} & = & \frac{c^{}_{\rm u}}{3} \left[ \left(\matrix{1 & 1
& 1 \cr 1 & 1 & 1 \cr 1 & 1 & 1}\right) + r^{}_{\rm u}
\left(\matrix{1 & 0 & 0 \cr 0 &
1 & 0 \cr 0 & 0 & 1}\right)\right] + \Delta M^{}_{\rm u} \; , \nonumber \\
M^{}_{\rm d} & = & \frac{c^{}_{\rm d}}{3}\left[ \left(\matrix{1 & 1
& 1 \cr 1 & 1 & 1 \cr 1 & 1 & 1}\right) + r^{}_{\rm d}
\left(\matrix{1 & 0 & 0 \cr 0 & 1 & 0 \cr 0 & 0 & 1}\right) \right]
+ \Delta M^{}_{\rm d} \; ,
\end{eqnarray}
where $c^{}_x >0$, $r^{}_x$ is real and $|r^{}_x| \ll 1$ holds (for
$x = {\rm u}$ and ${\rm d}$). For the sake of simplicity, the
textures of $\Delta M^{}_{\rm u}$ and $\Delta M^{}_{\rm d}$ are
taken to be exactly parallel to those of $\Delta M^{}_\nu$ and
$\Delta M^{}_l$ given in Eq. (8). In other words,
\begin{eqnarray}
\Delta M^{}_{\rm u} \; = \; \frac{c^{}_{\rm u}}{3} \left(\matrix{-
\delta^{}_{\rm u} & 0 & 0 \cr 0 & + \delta^{}_{\rm u} & 0 \cr 0 & 0
& \varepsilon^{}_{\rm u}}\right) \; , ~~~
\Delta M^{}_{\rm d} \; = \; \frac{c^{}_{\rm d}}{3}
\left(\matrix{-i\delta^{}_{\rm d} & 0 & 0 \cr 0 & +i\delta^{}_{\rm
d} & 0 \cr 0 & 0 & \varepsilon^{}_{\rm d}}\right) \;
\end{eqnarray}
with $0 < \delta^{}_x \ll \varepsilon^{}_x \ll 1$ (for $x = {\rm u}$
and ${\rm d}$). The diagonalization of $M^{}_{\rm u}$ and $M^{}_{\rm
d}$ is almost the same as that of $M^{}_l$. As a result,
\begin{eqnarray}
m^{}_t & = & c^{}_{\rm u} \left(1 +
\frac{\varepsilon^{}_{\rm u}}{9} + \frac{r^{}_{\rm u}}{3}\right) \; , ~~~
m^{}_c \; = \; c^{}_{\rm u} \left(\frac{2\varepsilon^{}_{\rm u}}{9} +
\frac{r^{}_{\rm u}}{3}\right) \; , ~~~
m^{}_u \; = \; c^{}_{\rm u} \left|\frac{r^{}_{\rm u}}{3} -
\frac{\delta^2_{\rm u}}{6\varepsilon^{}_{\rm u}} \right| \; ; ~~~~
\nonumber \\
m^{}_b & = & c^{}_{\rm d} \left(1 +
\frac{\varepsilon^{}_{\rm d}}{9} + \frac{r^{}_{\rm d}}{3}\right) \; , ~~~
m^{}_s \; = \; c^{}_{\rm d} \left(\frac{2\varepsilon^{}_{\rm d}}{9} +
\frac{r^{}_{\rm d}}{3}\right) \; , ~~~
m^{}_d \; = \; c^{}_{\rm d} \left|\frac{r^{}_{\rm d}}{3} +
\frac{\delta^2_{\rm d}}{6\varepsilon^{}_{\rm d}} \right| \; .
\end{eqnarray}
Defining $m^\prime_0 = c^{}_{\rm u} r^{}_{\rm u}/3$ and
$m^{\prime\prime}_0 = c^{}_{\rm d} r^{}_{\rm d}/3$, we have
\begin{eqnarray}
\varepsilon^{}_{\rm u} & = & \frac{9}{2} \frac{m^{}_c -
m^\prime_0}{m^{}_t - m^\prime_0} \; , ~~~ \delta^{}_{\rm u} \;
\approx \; \frac{2\varepsilon^{}_{\rm u}}{\sqrt{3}}
\frac{\sqrt{|m^{}_u - |m^\prime_0||}}{\sqrt{m^{}_c - m^\prime_0}} \;
,\nonumber \\
\varepsilon^{}_{\rm d} & = & \frac{9}{2} \frac{m^{}_s - m^{\prime
\prime}_0}{m^{}_b - m^{\prime \prime}_0} \; , ~~~ \delta^{}_{\rm d}
\; \approx \; \frac{2\varepsilon^{}_{\rm d}}{\sqrt{3}}
\frac{\sqrt{|m^{}_d - |m^{\prime \prime}_0||}}{\sqrt{m^{}_s -
m^{\prime \prime}_0}} \; ,
\end{eqnarray}
where the phenomenologically uninteresting case with $0 < m^\prime_0
< m^{}_u$ has been discarded. Furthermore, Eq. (20) gives rise to
the constraints $|m^\prime_0| < m^{}_u$ for $m^\prime_0 < 0$ and
$m^{\prime \prime}_0 < m^{}_d$ for $m^{\prime \prime}_0 > 0$. Given
$V^\dagger_{\rm u} M^{}_{\rm u} V^*_{\rm u} = {\rm Diag}\{m^{}_u,
m^{}_c, m^{}_t \}$ and $V^\dagger_{\rm d} M^{}_{\rm d} V^*_{\rm d} =
{\rm Diag}\{m^{}_d, m^{}_s, m^{}_b \}$, a careful perturbation
calculation yields
\begin{eqnarray}
V^{}_{\rm u} & \approx & V^{}_0 + \frac{x^{}_{\rm u}}{\sqrt{6}}
\left(\matrix{ 1 & -\sqrt{3} & 0 \cr 1 & \sqrt{3} & 0 \cr -2 & 0 &
0} \right) + \frac{y^{}_{\rm u}}{2\sqrt{3}} \left(\matrix{ 0 &
\sqrt{2} & -1 \cr 0 & \sqrt{2} & -1 \cr 0 & \sqrt{2} &
2} \right)
\nonumber \\
&& - \frac{x^2_{\rm u}}{2\sqrt{6}} \left(\matrix{ \sqrt{3} & 1 & 0
\cr -\sqrt{3} & 1 & 0 \cr 0 & -2 & 0} \right) + \frac{x^{}_{\rm u}
y^{}_{\rm u}}{\sqrt{6}} \left(\matrix{ 3 & 0 & -\sqrt{6} \cr 3 & 0 &
\sqrt{6} \cr 3 & 0 & 0} \right) \; ,
\nonumber \\
V^{}_{\rm d} & \approx & V^{}_0 - \frac{i x^{}_{\rm d}}{\sqrt{6}}
\left(\matrix{ 1 & \sqrt{3} & 0 \cr 1 & -\sqrt{3} & 0 \cr -2 & 0 &
0} \right) + \frac{y^{}_{\rm d}}{2\sqrt{3}} \left(\matrix{ 0 &
\sqrt{2} & -1 \cr 0 & \sqrt{2} & -1 \cr 0 & \sqrt{2} & 2} \right)
\nonumber \\
&& - \frac{x^2_{\rm d}}{2\sqrt{6}} \left(\matrix{ \sqrt{3} & 1 & 0
\cr -\sqrt{3} & 1 & 0 \cr 0 & -2 & 0} \right) - \frac{ix^{}_{\rm d}
y^{}_{\rm d}}{\sqrt{6}} \left(\matrix{ 3 & 0 & \sqrt{6} \cr 3 & 0 &
-\sqrt{6} \cr 3 & 0 & 0} \right) \; ,
\end{eqnarray}
where $x^{}_{\rm q} \equiv \sqrt{3} \delta^{}_{\rm q}/(2
\varepsilon^{}_{\rm q})$ and $y^{}_{\rm q} \equiv 2
\varepsilon^{}_{\rm q}/9$ (for q = u or d) have been defined. The
quark flavor mixing matrix $V^{}_{\rm CKM} \equiv V^\dagger_{\rm u}
V^{}_{\rm d}$ \cite{CKM} turns out to be
\begin{eqnarray}
V^{}_{\rm CKM} \; \approx \; \left(\matrix{ 1 & x^{}_{\rm u} - i
x^{}_{\rm d} & \displaystyle -\frac{y^{}_{\rm d}\left(2ix^{}_{\rm d}
+ x^{}_{\rm u}\right) - 3 x^{}_{\rm u} y^{}_{\rm u}}{\sqrt{2}} \cr
-x^{}_{\rm u} - ix^{}_{\rm d} & 1 & \displaystyle \frac{y^{}_{\rm u}
- y^{}_{\rm d}}{\sqrt{2}} \cr \displaystyle \frac{ix^{}_{\rm
d}(y^{}_{\rm u} - 3y^{}_{\rm d}) - 2x^{}_{\rm u} y^{}_{\rm
u}}{\sqrt{2}} & \displaystyle -\frac{y^{}_{\rm u} - y^{}_{\rm
d}}{\sqrt{2}} & 1} \right) \; ,
\end{eqnarray}
where the ${\cal O}(x^2_{\rm u})$, ${\cal O}(x^2_{\rm d})$, ${\cal
O}(x^{}_{\rm u} x^{}_{\rm d})$ and ${\cal O}(y^{}_{\rm u} y^{}_{\rm
d})$ corrections to the diagonal elements of $V^{}_{\rm CKM}$ have
been neglected. This result clearly shows that the dominant term of
$V^{}_{\rm CKM}$ is the identity matrix $I$, and thus its three
flavor mixing angles must be small and sensitive to the values of
$m^\prime_0$ and $m^{\prime\prime}_0$. The latter should not be
ignored not only because they have an impact on $V^{}_{\rm CKM}$ but
also because they come from an intrinsic $S^{}_3$ symmetry term.

To examine whether Eq. (23) is compatible with current experimental
data or not, we make use of the following values of quark masses at
the electroweak scale \cite{XZZ08}:
\begin{eqnarray}
m^{}_u & = & \left(0.85 \sim 1.77\right)~{\rm MeV} \;, ~~\;\;\;~
m^{}_d \; = \; \left(1.71 \sim 4.14\right)~{\rm MeV} \; , \nonumber \\
m^{}_c & = & \left(0.535\sim 0.703\right)~{\rm GeV} \; , ~~~ m^{}_s
\; = \; \left(40 \sim 71\right)~{\rm MeV} \; , \nonumber \\
m^{}_t & = & \left(168.7 \sim 174.7\right)~{\rm GeV} \; , ~~~ m^{}_b
\; = \; \left(2.80\sim 2.98\right)~{\rm GeV} \; .
\end{eqnarray}
One may roughly take the geometric relations $m^{}_u/m^{}_c \approx
m^{}_c/m^{}_t \approx 1/300$ and $m^{}_d/m^{}_s \approx
m^{}_s/m^{}_b \approx 1/40$ to illustrate the strong quark mass
hierarchies. In our discussions $m^\prime_0 \ll m^{}_c$ and
$m^{\prime \prime}_0 \ll m^{}_s$ are reasonably assumed to assure
that the perturbation calculations are valid. So the magnitudes of
$\delta^{}_{\rm u}$ and $\delta^{}_{\rm d}$ are sensitive to
$m^\prime_0$ and $m^{\prime \prime}_0$, whereas those of
$\varepsilon^{}_{\rm u}$ and $\varepsilon^{}_{\rm d}$ are not. In
other words, $x^{}_{\rm u}$ (or $x^{}_{\rm d}$) may significantly
deviate from $\sqrt{m^{}_u/m^{}_c}$ (or $\sqrt{m^{}_d/m^{}_s}$)
while $y^{}_{\rm u}$ (or $y^{}_{\rm d}$) is essentially equal to
$m^{}_c/m^{}_t$ (or $m^{}_s/m^{}_b$). These observations are helpful
when we confront Eq. (23) with the experimental constraints on
$V^{}_{\rm CKM}$ \cite{PDG},
\begin{eqnarray}
|V^{}_{\rm CKM}| & = & \left(\matrix{|V^{}_{ud}| & |V^{}_{us}| &
|V^{}_{ub}| \cr |V^{}_{cd}| & |V^{}_{cs}| & |V^{}_{cb}| \cr
|V^{}_{td}| & |V^{}_{ts}| & |V^{}_{tb}| }\right)
\nonumber \\
& = & \left(\matrix{0.97419 \pm 0.00022 & 0.2257 \pm 0.0010 &
0.00359 \pm 0.00016 \cr 0.2256 \pm 0.0010 & 0.97334 \pm 0.00023 &
0.0415^{+0.0010}_{-0.0011} \cr 0.00874^{+0.00026}_{-0.00037} &
0.0407 \pm 0.0010 & 0.999133^{+0.000044}_{-0.000043}}\right) \; .
~~~
\end{eqnarray}
Some discussions are in order.
\begin{itemize}
\item      We obtain $|V^{}_{cb}| \approx |V^{}_{ts}| \approx
|y^{}_{\rm u} - y^{}_{\rm d}|/\sqrt{2} \approx (m^{}_s/m^{}_b -
m^{}_c/m^{}_t)/\sqrt{2}$ in the leading-order approximation, which
is too low to fit the observed values of $|V^{}_{cb}|$ and
$|V^{}_{ts}|$. The reason for this discrepancy is simple: the
diagonal patterns of $\Delta M^{}_{\rm u}$ and $\Delta M^{}_{\rm d}$
taken in Eq. (19) fail in producing a sufficiently large mixing
angle between the second and third families in the quark sector. In
fact, a similar problem has appeared in the lepton sector: the
diagonal perturbation matrices $\Delta M^{}_l$ and $\Delta M^{}_\nu$
in Eq. (8) are unable to produce $\theta^{}_{23} \approx \pi/4$ for
the atmospheric neutrino mixing angle. A straightforward way out
should be to find out a class of different patterns of $\Delta
M^{}_x$ (for $x = l, \nu, {\rm u}, {\rm d}$), which might be
off-diagonal or partially off-diagonal but can simultaneously make
the mixing angle $\theta^{}_{23}$ of $V^{}_{\rm MNS}$ and the matrix
elements $|V^{}_{cb}|$ and $|V^{}_{ts}|$ of $V^{}_{\rm CKM}$ large
enough \cite{XYZ}.

\item      Here we are concerned about whether it is possible to
achieve the experimentally-favored values of
$|V^{}_{ub}|/|V^{}_{cb}|$ and $|V^{}_{td}|/|V^{}_{ts}|$ in the
presence of non-vanishing $m^\prime_0$ and $m^{\prime\prime}_0$,
because many of the hitherto-proposed quark mass ans$\rm\ddot{a}$tze
predict $|V^{}_{ub}|/|V^{}_{cb}| \approx \sqrt{m^{}_u/m^{}_c}$
\cite{Du} which is badly lower than the present experimental value
of $|V^{}_{ub}|/|V^{}_{cb}|$. Eq. (23) leads us to
\begin{eqnarray}
\frac{|V^{}_{ub}|}{|V^{}_{cb}|} = \frac{\sqrt{4 x^2_{\rm d} y^2_{\rm
d} + x^2_{\rm u}(y^{}_{\rm d} - 3y^{}_{\rm u})^2}}{|y^{}_{\rm u} -
y^{}_{\rm d}|} \; , ~~~ \frac{|V^{}_{td}|}{|V^{}_{ts}|} =
\frac{\sqrt{4 x^2_{\rm u} y^2_{\rm u} + x^2_{\rm d}(y^{}_{\rm u} -
3y^{}_{\rm d})^2}}{|y^{}_{\rm u} - y^{}_{\rm d}|} \; .
\end{eqnarray}
In the $m^\prime_0 = m^{\prime\prime}_0 \approx 0$ case, we simply
obtain $|V^{}_{ub}|/|V^{}_{cb}| \approx 2\sqrt{m^{}_d/m^{}_s}$ and
$|V^{}_{td}|/|V^{}_{ts}| \approx 3\sqrt{m^{}_d/m^{}_s}$ from Eq.
(26). These two results are apparently in conflict with the
experimental data given in Eq. (25). Hence one has to switch on the
contributions from $m^\prime_0$ and $m^{\prime\prime}_0$. To see how
large $m^\prime_0$ and $m^{\prime\prime}_0$ should be, we rewrite
Eq. (26) as
\begin{eqnarray}
x^2_{\rm u} & = & \frac{1}{3 \left(3R^2 - 14 R + 3\right)} \left[
\left(R - 3\right)^2 \frac{|V^{}_{ub}|^2}{|V^{}_{cb}|^2} - 4
\frac{|V^{}_{td}|^2}{|V^{}_{ts}|^2}\right] \; ,
\nonumber \\
x^2_{\rm d} & = & \frac{1}{3 \left(3R^2 - 14 R + 3\right)} \left[
\left(1 - 3R\right)^2 \frac{|V^{}_{td}|^2}{|V^{}_{ts}|^2} - 4 R^2
\frac{|V^{}_{ub}|^2}{|V^{}_{cb}|^2} \right] \; ,
\end{eqnarray}
where $R \equiv y^{}_{\rm u}/y^{}_{\rm d}$. Because $y^{}_{\rm u}$
and $y^{}_{\rm d}$ are almost insensitive to $m^\prime_0$ and
$m^{\prime\prime}_0$, we approximately have $R \approx (m^{}_c
m^{}_b)/(m^{}_t m^{}_s) \in [0.12, ~0.31]$ from Eq. (24). Typically
taking $R \approx 0.3$ and adopting the best-fit values of
$|V^{}_{ub}|/|V^{}_{cb}|$ and $|V^{}_{td}|/|V^{}_{ts}|$ given in Eq.
(25), we find $x^2_{\rm u} \approx 4.6 \times 10^{-2}$ and $x^2_{\rm
d} \approx  8.0 \times 10^{-4}$. Accordingly, we get $||m^\prime_0|
- m^{}_u| \approx m^{}_c x^2_{\rm u} \approx 14 m^{}_u$ and
$||m^{\prime\prime}_0| - m^{}_d| \approx m^{}_s x^2_{\rm d} \approx
3.2 \times 10^{-2} m^{}_d$ if we simply input $m^{}_u/m^{}_c \approx
1/300$ and $m^{}_d/m^{}_s \approx 1/40$. In view of $r^{}_{\rm u} =
3m^\prime_0/c^{}_{\rm u} \approx 3m^\prime_0/m^{}_t$ and $r^{}_{\rm
d} = 3m^{\prime\prime}_0/c^{}_{\rm d} \approx
3m^{\prime\prime}_0/m^{}_b$, we obtain $r^{}_{\rm u} \approx 45
m^{}_u/m^{}_t \approx 5\times 10^{-4}$ (for $m^\prime_0 > m^{}_u$)
as well as $r^{}_{\rm d} \approx \pm 3m^{}_d/m^{}_b \approx \pm
1.9\times 10^{-3}$ (for $|m^{\prime\prime}_0| \sim m^{}_d$). The
smallness of both $r^{}_{\rm u}$ and $r^{}_{\rm d}$ is consistent
with the strong mass hierarchies of up- and down-type quarks which
have been described by the democratic $D$ term in the $S^{}_3$
flavor symmetry limit.

\item     In this ansatz the Cabibbo angle $\theta^{}_{\rm C}$
(i.e., $\sin\theta^{}_{\rm C} \equiv |V^{}_{us}| \approx
|V^{}_{cd}|$) is given by
\begin{eqnarray}
\theta^{}_{\rm C} \; =\; \arcsin \left(|V^{}_{us}|\right) \; \approx
\; \arcsin \left(|V^{}_{cd}|\right) \; \approx \;
\arcsin\left(\sqrt{x^2_{\rm u} + x^2_{\rm d}} \right) \; .
\end{eqnarray}
With the inputs $x^2_{\rm u} \approx 4.6 \times 10^{-2}$ and
$x^2_{\rm d} \approx  8.0 \times 10^{-4}$ obtained above, we obtain
$\theta^{}_{\rm C} \approx 12.5^\circ$ or equivalently $|V^{}_{us}|
\approx |V^{}_{cd}| \approx 0.22$. Such a result is quantitatively
compatible with current experimental data, but it is qualitatively
different from the conventional prediction $\sin\theta^{}_{\rm C}
\approx \sqrt{m^{}_d/m^{}_s}$ or $\sin\theta^{}_{\rm C} \approx
|\sqrt{m^{}_d/m^{}_s} - e^{i\phi} \sqrt{m^{}_u/m^{}_c}|$ with $\phi$
being a free phase parameter based on a class of quark mass
ans$\rm\ddot{a}$tze \cite{Cabibbo1,Cabibbo2,Cabibbo3}.

\item     Since CP violation has been observed in the quark
sector, let us take a look at the consequences of our ansatz on four
well-known rephasing invariants of CP violation: the Jarlskog
parameter ${\cal J}$ \cite{J} and three inner angles of the CKM
unitarity triangle $V^{}_{ud} V^*_{ub} + V^{}_{cd} V^*_{cb} +
V^{}_{td} V^*_{tb} =0$ \cite{PDG}. In the leading-order
approximation, we obtain
\begin{equation}
{\cal J} \; \equiv \; {\rm Im} \left(V^{}_{us} V^{}_{cb} V^*_{ub}
V^*_{cs} \right) \; \approx \; -\frac{3}{2} x^{}_{\rm u} x^{}_{\rm
d} \left( y^{}_{\rm u} - y^{}_{\rm d} \right)^2 \; \approx \;
-3 x^{}_{\rm u} x^{}_{\rm d} |V^{}_{cb}|^2 \; , ~~
\end{equation}
and
\begin{eqnarray}
\alpha & \equiv & \arg \left(- \frac{V^{}_{td} V^*_{tb}}{V^{}_{ud}
V^*_{ub}} \right ) \; \approx \; \arctan \left[ \frac{\displaystyle
3 x^{}_{\rm u} x^{}_{\rm d} \left( y^{}_{\rm u} - y^{}_{\rm d} \right)^2}{\displaystyle 2 \left[ x^2_{\rm
u} y^{}_{\rm u} \left( y^{}_{\rm d} - 3 y^{}_{\rm u} \right) +
x^2_{\rm d} y^{}_{\rm d} \left( y^{}_{\rm u} - 3 y^{}_{\rm d}
\right) \right]} \right]
\; , \nonumber \\
\beta & \equiv & \arg \left(- \frac{V^{}_{cd} V^*_{cb}}{V^{}_{td}
V^*_{tb}} \right ) \; \approx \;
\arctan\left[ \frac{\displaystyle
3 x^{}_{\rm u} x^{}_{\rm d} \left( y^{}_{\rm u}
- y^{}_{\rm d} \right)}{\displaystyle 2 x^2_{\rm u} y^{}_{\rm u}
- x^2_{\rm d} \left( y^{}_{\rm u} - 3 y^{}_{\rm d} \right)}
\right]
\; , \nonumber \\
\gamma & \equiv & \arg \left(- \frac{V^{}_{ud} V^*_{ub}}{V^{}_{cd}
V^*_{cb}} \right ) \; \approx \;
\arctan\left[ \frac{\displaystyle 3 x^{}_{\rm u} x^{}_{\rm d}
\left( y^{}_{\rm u} - y^{}_{\rm d} \right)}
{\displaystyle x^2_{\rm u} \left( y^{}_{\rm d} - 3 y^{}_{\rm u}
\right) - 2x^2_{\rm d} y^{}_{\rm d}} \right]
\; .
\end{eqnarray}
Current experimental data yield $|V^{}_{cb}| \approx 0.0415$, and the
numerical exercises done above allow $R \equiv y^{}_{\rm u}/
y^{}_{\rm d} \approx 0.3$,
$x^2_{\rm u} \approx 4.6 \times 10^{-2}$ and $x^2_{\rm d} \approx 8.0
\times 10^{-4}$. Assuming $x^{}_{\rm u}$ and $x^{}_{\rm d}$ to
have the opposite sign, we immediately obtain ${\cal J} \approx 3.1 \times
10^{-5}$. This result is in good agreement with the value of ${\cal J}$
given in Ref. \cite{PDG}. In addition, we
estimate the values of $\alpha$, $\beta$
and $\gamma$ in Eq. (30) and arrive at $\alpha \approx 80^\circ$,
$\beta \approx 23^\circ$ and $\gamma \approx 77^\circ$, which are essentially
compatible with current experimental data \cite{PDG}.
\end{itemize}
Of course, our results depend on the diagonal perturbation patterns
of $\Delta M^{}_{\rm u}$ and $\Delta M^{}_{\rm d}$ taken in Eq.
(19). We shall carry out a systematic analysis of different forms of
$\Delta M^{}_x$ (for $x= l, \nu, {\rm u}, {\rm d}$) elsewhere
\cite{XYZ}, in order to find out simpler and more realistic textures
of fermion mass matrices motivated by the $S^{}_3$ flavor symmetry.

\vspace{0.3cm}

\framebox{\large\bf 4} ~ We have extended the standard electroweak
model by introducing one Higgs triplet and realizing the type-II
seesaw mechanism to generate finite neutrino masses, and then
applied the $S^{}_3$ flavor symmetry to both lepton and quark
sectors. The resulting mass matrices of charged leptons, neutrinos,
up-type quarks and down-type quarks have a universal form $M^{}_x =
\xi^{}_x I + \zeta^{}_x D$ with $I$ being the identity matrix and
$D$ being the democracy matrix (for $x= l, \nu, {\rm u}, {\rm d}$).
The observed mass hierarchies of charged fermions can therefore be
understood in the $S^{}_3$ symmetry limit with $\xi^{}_x \ll
\zeta^{}_x$ (for $x= l, {\rm u}, {\rm d}$). We have also argued that
$\xi^{}_\nu \gg \zeta^{}_\nu$ is likely to hold, implying a nearly
degenerate mass spectrum for three neutrinos. Such a picture
generally allows us to interpret why the quark flavor mixing matrix
is close to $I$ and why the lepton flavor mixing matrix may contain
two large mixing angles originating from the diagonalization of $D$,
after proper perturbations to $M^{}_l$, $M^{}_\nu$, $M^{}_{\rm u}$
and $M^{}_{\rm d}$ are taken into account. However, the $I$ terms of
$M^{}_l$, $M^{}_{\rm u}$ and $M^{}_{\rm d}$ used to be ignored or
dictated to be zero by means of the $S^{}_{3 \rm L} \times S^{}_{3
\rm R}$ symmetry. We have stressed that this term is important for
two reasons: on the one hand, it is intrinsical in the $S^{}_3$
symmetry limit and should not be forgotten; on the other hand, it
can significantly modify the smallest lepton flavor mixing angle
$\theta^{}_{13}$ or the off-diagonal elements of the quark flavor
mixing matrix. To illustrate, we have discussed a diagonal
perturbation ansatz of $M^{}_x$ to explicitly break the $S^{}_3$
symmetry and calculate the mass spectra and flavor mixing patterns
of leptons and quarks. We find that this simple ansatz is
phenomenologically acceptable in the lepton sector, but it is
difficult to produce correct $|V^{}_{cb}|$ and $|V^{}_{ts}|$ in the
quark sector.

We have pointed out that a straightforward way to modify our ansatz
is to consider different patterns of the perturbation matrices
$\Delta M^{}_l$, $\Delta M^{}_\nu$, $\Delta M^{}_{\rm u}$ and
$\Delta M^{}_{\rm d}$. This requires a systematic analysis of
fermion mass matrices based on the $S^{}_3$ flavor symmetry, because
there are many possibilities and the main criterion to select the
favorite one is to see that its phenomenological consequences
can be consistent very well with current
experimental data on lepton and quark masses and flavor mixing
parameters. We are going to do such a comprehensive but tedious
analysis elsewhere \cite{XYZ}.

One may also consider to combine the $S^{}_3$ flavor symmetry with
the type-I seesaw mechanism, the type-(I+II) seesaw mechanism or
other mechanisms of neutrino mass generation \cite{RX04,S3}. Of course,
it is more interesting to simultaneously understand the lepton and
quark flavor structures by means of the $S^{}_3$ symmetry and its
proper breaking mechanism. Some enlightening ideas in this kind of exercises
are expected to be very helpful in the study of other discrete flavor
symmetries and their consequences or implications on fermion mass generation,
flavor mixing and CP violation.

\vspace{0.3cm}

This work was supported in part by the National Natural Science Foundation
of China under grant No. 10425522 and No. 10875131 (Z.Z.X.) and No.
10705050 (D.Y.), in part by the Ministry of Science and Technology
of China under grant No. 2009CB825207 (Z.Z.X.), and in part by the
Alexander von Humboldt Foundation (S.Z.).

\end{document}